\documentclass[12pt]{article}
\usepackage{amssymb,amsmath,esint,titlesec,mathrsfs}
\titleformat*{\section}{\normalsize\bf}
\titleformat*{\subsection}{\small\bf}

\begin{document}


\begin{titlepage}

\setlength{\baselineskip}{18pt}

                               \vspace*{0mm}

                             \begin{center}

{\Large\bf  Composition of q-entropies and hyperbolic orthogonality }

                                   \vspace{45mm}

              \normalsize\sf  NIKOLAOS \  \  KALOGEROPOULOS $^\dag$\\

                            \vspace{1mm}

                   {\small\sf  Center for Research and Applications \\
                                  of Nonlinear Systems  \  \  (CRANS),\\
                          University of Patras, Patras 26500, Greece.\\ }
                                
                                      \end{center}

                            \vspace{35mm}

                     \centerline{\large\bf Abstract}
                                
                                     \vspace{3mm}
                     
    \noindent We point out that the q-entropy composition for independent events has exactly the same form as the 
    Pythagorean theorem in hyperbolic geometry. We justify the formal relation 
    of hyperbolic geometry with the q-entropy through the use of the 
    $\kappa$-entropy, which is directly related to the hyperboloid model of hyperbolic space.  
    We comment on the  relation between orthogonality in this form of the Pythagorean theorem 
    and the independence of the probability 
    distributions appearing in the q-entropy composition through the use of the Dvoretzky-Rogers lemma.\\

                                             \vfill


                             \vspace{0mm}

\noindent\rule{12.5cm}{0.3mm}\\  
   \noindent   {\small\rm $^\dag$        Electronic mail: \ \ \   {\sf nikos.physikos@gmail.com}} \\
   
\end{titlepage}
 

                                                                                \newpage                 

\rm\normalsize
\setlength{\baselineskip}{18pt}

\section{Introduction} 

For more than three decades, the q-entropy (also called ``Tsallis entropy'') \cite{Tsallis}, which is a one-parameter family of entropies, 
has been a subject of investigation in Statistical Mechanics. Despite some progress, it is probably fair to say that its dynamical foundations 
and domain of applicability in Statistical Physics, if any, are still shrouded in mystery. As a result, speculation abounds, and people have turned
mostly to data fittings in an attempt to justify the use of q-entropy, primarily through its extremal distributions under some constraints, 
the q-exponentials \cite{Tsallis}.  One would, ideally, like to have a microscopic basis for the use  of the q-entropy 
which would crucially differentiate it from the well-established Boltzmann/Gibbs/Shannon (BGS)  entropy. \\

A way to determine differences between the BGS and the q-entropy is to compare their axiomatic formulations \cite{Santos}, \cite{Abe1}, \cite{Abe2}, 
\cite{ET}. From this viewpoint, the major difference, at the axiomatic level,  between the q- and the BGS entropy is in their addition/composition property for 
probabilistically independent occurrencies \cite{Tsallis}.   \\

Several authors have used, in elucidating aspects of q-entropy, this axiomatic approach as starting point, or the more general but related approach 
based on the  algebraic properties of composition of entropies, as for instance \cite{T1}, \cite{T2}, \cite{JT1}, \cite{JT2}.  Our investigation is somewhat related, but 
quite different in that it relies heavily on geometric, rather than on algebraic, structures related to the q-entropy in order to elucidate its composition law. 
We follow a different path from information geometry \cite{Amari}, as we do not rely on information theory or statistics in any way,  
even though one can clearly discern  some rare similarites between our treatments.\\

We point out that there is a formal analogy between the Pythagorean theorem in hyperbolic space and the q-entropy composition for independent 
occurrencies.  We trace back the origin of such a  ``hyperbolic behavior'' in the composition of the
q-entropy to the relation between the q-entropy and the $\kappa$-entropy \cite{WS}. 
This latter entropy has been motivated by and  relies on properties of the Lorentz transformations 
\cite{Kan1}, \cite{Kan2}, \cite{Kan3}, \cite{Kan4} which are in turn related to hyperbolic spaces
through the hyperboloid model of hyperbolic geometry. \\

In Section 2, we derive the expression of the Pythagorean theorem for hyperbolic spaces which we need in this work.
In Section 3, we comment on  the relation between the q- and the $\kappa$- entropies which justify the hyperbolic nature of the q-entropy composition law.
In Section 4, we present formal statements mainly from Convex Geometry, which attempt to justify the relation between
orthogonality and probabilistic independence. 
Section 5  contains further comments related to this work and a general outlook toward independence and additivity. \\      


\section{The Pythagorean theorem for hyperbolic spaces}  

There is little doubt that the Pythagorean theorem is probably the most important metric relation in all of Euclidean Geometry.
It forms the basis for its extensions in numerous directions. The most relevant for Physics have been the quadratic forms determining 
infinitesimal distances in Riemannian and semi-Riemannian geometry, the former of which is used in systems requiring positive-definite metrics 
where the latter are usually employed in relation to Special or General Relativity or theories which are based on or extend their formalism.\\

What is probably less known is that the Pythagorean theorem has an analogue in hyperbolic geometry. This is somewhat surprising as most people tend to think 
about the Pythagorean theorem  as the staple of Euclidean metrics, something however which does not preclude a similar relation to hold for other geometries.  
From the viewpoint of Riemannian geometry, the hyperbolic space is the unique simply connected $n$-dimensional manifold with constant sectional curvature $-1$.
It should be mentiorned that hyperbolic geometry arose in the early 19th century in a synthetic/axiomatic fashion as an attempt to determine whether Euclid's 
fifth axiom was independent or could be derived from the first four, a quest which lasted almost two millenia \cite{Th}. One can appreciate the strong visualization
and excellent geometric intuition provided about hyperbolic geometry when it is seen in its axiomatic/synthetic form. The synthetic approach is realized
by employing one  of the several well-known models of hyperbolic space \cite{CFKP}. \\     

We will not attempt to provide a proof of the hyperbolic form of the Pythagorean theorem here, since it involves a relatively long process which even though is very instructive from a geometric viepoint, it offers very little for our purposes. Moreover  such a proof can easily be found in the print and electronic literature.  
Instead we are content to start from the conventionally accepted form of the hyperbolic Pythagorean theorem which states that 
for a hyperbolic triangle ABC, with the right angle being at C,\
\begin{equation} 
   \cosh (c) = \cosh (a) \cosh (b)
\end{equation}
where \ $a$ \ is the length of the side opposite to angle \ $A$, \ and similarly for \ $b$ \ and \ $c$. \ Squaring both sides and using the obvious identity
\begin{equation}  
     \cosh^2(x) = \sinh^2(x) + 1
\end{equation}
after a simplification, we get 
\begin{equation}
         \sinh^2(c) = \sinh^2(a) + \sinh^2(b) + \sinh^2(a)\sinh^2(b)
\end{equation}
Setting \ $d_a = \sinh (a), \ d_b = \sinh (b), \ d_c = \sinh (c)$, \ we finally get
\begin{equation} 
  d_c^2 = d_a^2 + d_b^2 +d_a^2 d_b^2
\end{equation}
which is the sought-after form of the Pythagorean theorem in hyperbolic space which we will be using in the rest of this work. Upon closer 
inspection it will turn out that (4) has the same form as the q-entropy composition property for independent occurrences 
as will be pointed out in the next Section.\\ 


\section{Hyperbolicity of q-entropies through $\kappa$-entropies} 

The q-entropy (also known as ``Tsallis entropy'') \cite{Tsallis} for a discrete sample space \ $\mathfrak{X}$ \ whose elements are indexed by a set \ $I$ \ with  probabilities 
\ $p_i, \ i\in I$ \ is defined as 
\begin{equation}
           \mathcal{S}_q [\{ p_i \} ] \  = \   k_B \cdot \frac{1}{q-1} \left( 1 - \sum_{i\in I} \  p_i^q \right)
\end{equation}
where \ $k_B$ \ stands for the Boltzmann constant to be set to one for convenience, 
and where \ $q\in\mathbb{R}$ \ is a constant called the ``entropic parameter''. 
In practice \ $q$ \ is usually  restricted to a proper subset of \ $\mathbb{R}$ \ in most cases of physical interest.   
One immediately sees that for two independent sets of occurrences \ $A, B \subset \mathfrak{X}$, \ namely if 
\begin{equation}  
     p_{A\cap B} = p_A \cdot p_B
\end{equation}
then 
\begin{equation}
      \mathcal{S}_q [ p_{A\cap B} ] = \mathcal{S}_q [p_A] + \mathcal{S}_q [p_B] + (1-q) \mathcal{S}_q [p_A] \mathcal{S}_q [p_B] 
\end{equation}
and defining 
\begin{equation} 
      \mathcal{S}^\prime _q = \frac{1}{1-q} \ \mathcal{S}_q
\end{equation}
the composition property (7) becomes 
\begin{equation}
      \mathcal{S}^\prime _q [ p_{A\cap B} ] = \mathcal{S}^\prime _q [p_A] + \mathcal{S}^\prime _q [p_B] + \mathcal{S}^\prime _q [p_A] \mathcal{S}^\prime _q [p_B] 
\end{equation}
which has exactly the same form as the Pythagorean theorem in hyperbolic space (4).\\

One question which naturally arises is whether the q-entropy composition property (9) has anything of substance to do with the Pythagorean theorem in hyperbolic space, except for their formal similarity. This question can be further broken down to two, largely indepedent of each other, questions
\begin{enumerate}   
  \item What does the q-entropy have to do with hyperbolic space? 
  \item What does orthogonality have to do with the composition property (9)? 
\end{enumerate}
It seems that the fact that the q-entropy composition has a formal similarity to Pythagorean theorem in hyperbolic space has been pointed out before in the context
of information geometry  \cite{Amari}. But so far as we know there has not been any attempt to understand \emph{why} the composition property (9) has the same form as the hyperbolic version of the  Pythagorean theorem (4), and instead it has been  taken at face value. 
The information geometry practitioners focused instead on creating and further developing a rather elaborate formalism of dual 
Riemannian geometric structures and escort distributions \cite{Amari}. 
Therefore the present work may help address this gap in our understanding of q-entropy, in our opinion. \\         

We begin by stating that the q-entropy is not a distance function as one might be tempted to assume at first glance. This is indeed true for most entropic forms
which have been proposed over the years \cite{IKGS}, including the BGS entropy, despite occasional comments in the literature about the opposite.
Usually such entropic forms fail to obey the triangle inequality. 
Moreover one sees such entropies as ways of measuring the difference between two probability distributions, in order to establish  coordinate reparametrization invariance, 
akin to the Kullback-Leibler, the Bregman etc  divergences. 
Then such entropic forms fail to obey even the symmetry property in their arguments, which is required of distance functions. 
Hence, one should be careful not to take the analogy too far or draw any invalid conclusions by the formal similarity between (4) and (9).\\
     
The answer to question 1 comes through the relation of the q- and the $\kappa$- entropies. 
The $\kappa$-entropy \ $\mathcal{R}_\kappa [p_i]$ \ is a one-parameter set of entropies defined 
\cite{Kan1}, \cite{Kan2}, \cite{Kan3}, \cite{Kan4} for discrete sets of occurrences \ $p_i, \ i\in I $ \ by 
\begin{equation}
  \mathcal{R}_\kappa [\{ p_i \} ] \ = \ k_B \sum_{i\in I} \  \frac{p_i^{1-\kappa} - p_i^{1+\kappa}}{2\kappa}
\end{equation}
The $\kappa$-entropy purports to generalize the BGS entropy to a special relativistic context and is partially motivated by the $\kappa$ distributions 
which have been observed in the heliosphere, and which have been a subject of intense investigation over several decades \cite{Cook} especially in plasma 
Astrophysics. \\

At this point, and for comparison purposes to the q-entropy composition (9) only, we digress to state that the composition property of the $\kappa$-entropy 
for independent probabilities is not as easy to obtain as the composition of q-entropies.  It was eventually found to be \cite{WSSW} 
\begin{equation}
     \mathcal{R}_\kappa [p_{A\cap B}] \  =  \ \mathcal{R}_\kappa [p_A] \left( \frac{1}{\gamma_\kappa} \mathcal{R}_\kappa \left[ p_B/\eta_\kappa \right] - 
                                                                                                                          \mathcal{R}_\kappa [p_B] \right)
                                                                   + \mathcal{R}_\kappa [p_B] \left( \frac{1}{\gamma_\kappa} \mathcal{R}_\kappa \left[ p_A/\eta_\kappa \right] - 
                                                                                                                             \mathcal{R}_\kappa [p_A] \right) 
\end{equation}
where 
\begin{equation}
      \gamma_\kappa = \frac{1}{\sqrt{1-\kappa^2}}
\end{equation}
is a rapidity-like factor, and 
\begin{equation}
       \eta_\kappa = \left(  \frac{1+\kappa}{1-\kappa} \right)^\frac{1}{2\kappa}
\end{equation}
The $\kappa$-entropy has a functional form which is directly descended from the Lorentz transformations in Special Relativity \cite{Kan1},
\cite{Kan2}. Since the Lorentz group is a normal subgroup of the Poincar\'{e} group, which is the group of isometries of Minkowski space-time, following the 
viewpoint  of F. Klein's Erlangen Programme, one can see that properties of the $\kappa$-entropy are formally related to properties of 
Minkowski space-time. \\

The relation between the q- and $\kappa$-entropies is readily seen to be 
\begin{equation}
          \mathcal{R}_\kappa [p_i] = \frac{1}{2} \left( \mathcal{S}_{1+\kappa} [p_i] + \mathcal{S}_{1-\kappa} [p_i] \right)
\end{equation} 
This relation can be inverted to give the q-entropies in terms of the $\kappa$-entropies, but the explicit form is not necessary for our argument.
All that matters is that there is a simple relation between these two entropic forms. As was mentioned before, the $\kappa$-entropy reflects properties of 
Minkowski space-time. To be more concrete, let
\ $(x_1, x_2, \ldots, x_{n+1}) \in \mathbb{R}^{n+1}$ \ endowed with the quadratic form    
\begin{equation}
    ds^2 = dx_1^2+ dx^2_2+ \ldots +dx_n^2 - dx_{n+1}^2
\end{equation}
Consider the $n$-dimensional hypersurface defined as the ``sphere of unit imaginary radius'' namely
\begin{equation}
      \mathbb{H}^n = \{ (x_1, x_2, \ldots, x_n, x_{n+1})\in \mathbb{R}^{n+1}: \ \ x_1^2+x_2^2+\ldots x_n^2- x_{n+1}^2 = -1, \ \ x_{n+1} >0 \}  
\end{equation}   
This  single branch hyperboloid has a space-like, namely a Riemannian, metric  and a careful examination of its properties proves that it is actually isometric to the 
$n$-dimensional hyperbolic space \ $\mathbb{H}^n$, \ hence the symbol we used for it in (16). 
This is actually the well-known hyperboloid model \cite{Th}, \cite{CFKP} of hyperbolic space. \\

To finish the argument, the hyperbolic space appears as a subset of Minkowski 
space-time, the latter of  which has some common properties, formaly of course, with the $\kappa$-entropy. 
On the other hand, the q-entropy is a combination of $\kappa$-entropies. 
Therefore, it is not totally surprising that some of the formal properties of the hyperbolic space are inherited by the q-entropy. 
The composition property of the q-entropy is one of them. 
This justifies the formal resemblance of the hyperbolic Pythagorean theorem (4) and the q-entropy composition 
property (9), so far as the word ``hyperbolic'' is concerned. The justification of the word ``orthogonal'' will be dealt with in the next Section.\\

Before finishing this Section, it may worth making a couple of comments. In some of our prior works \cite{NK}, \cite{CK} we noticed that the 
formal resemblance  between the q-entropy composition property and elements of sub-Riemannian geometry related to the Heisenberg, or more generally, 
to nilpotent groups. The problem with Carnot groups, or more generally Carnot-Carath\'{e}odory spaces \cite{Gr2} 
is that they have a lot of structure, so in a sense they are very special. On the other hand, manifolds or more general spaces of negative curvature are plentiful.      
An example of this is dramatically expressed by W. Thurston's work on geometrization \cite{Th} which showed 
that among all 3-dimensional manifolds most of them are hyperbolic. Let us not forget that something similar occurs with surfaces (manifolds of dimension 2) 
since surfaces of any genus greater than one admit a hyperbolic metric. But these are low-dimensional results which are not of great interest if one works with
configuration or phase spaces of systems with many degrees of freedom, which are of main interest in Statistical Mechanics.  \\

A more pertinent result in the use of hyperbolic structures valid for  manifolds of high dimension though can be found in \cite{Ont} where it is proved that a  
closed smooth manifold of any dimension can be given a metric of negatively pinched curvature close to a hyperbolic one, even without singularities.
This when combined with the Mostow rigidity, which states that two hyperbolic manifolds of dimension greater than two which have isomorphic 
fundamental groups are actually isometric \cite{Mostow}, makes us suspect that hyperbolicity in high dimensional manifolds may be more frequent than one might have naively thought. \\
     
The result of all this is that  from the viewpoint of geometry, the composition of q-entropies which has the same form as the hyperbolic Pythagorean theorem 
may be quite a bit more frequent than one might believe at a first glance. 
Since this composition is the distinguishing property between the BGS and q-entropies, the latter may be applicable to more physical 
systems than might have initially been suspected. \\ 
  

\section{Orthogonality and probabilistic independence}

The 	question we wish to address in this Section, is question 2 in the list of the previous Section, namely, what  is the reason for the formal orthogonality between the joint probabilty 
distribution and its marginals appearing in the composition law (9). 
Since the standard hyperbolic metric and the Euclidean one are conformal \cite{Th}, \cite{CFKP} the angles between lines in the hyperbolic plane \ $\mathbb{H}^2$ 
\ and in the Euclidean plane \ $\mathbb{R}^2$ \ are equal. So  it is sufficient to make an argument for orthogonality for probability distributions 
in \ $\mathbb{R}^2$. \\  

Let us consider as sample space a finite set \ $\mathfrak{X} = \{ a_1, a_2, a_3, \ldots, a_n \} $ \ in order to avoid  details related to infinities and to continuous distributions that would obscure the argument by making  it far more technical but would add nothing of essence.  
Assign probabilities \ $p_1, \ p_2, \ p_3, \ldots$ \ to each element \ $a_1, \ a_2, \ a_3, \ldots, a_n$ \ of \ 
$\mathfrak{X}$ \ respectively. \ Obviously
\begin{equation}
      \sum_{i=1}^n \ p_i = 1,  \ \ \ p_i \geq 0
\end{equation}
These probabilities form the vertices of a probability simplex \ $\Delta^{n-1} \subset \mathbb{R}^n$ \ when the value of each one of them is placed 
on the axes of an orthonormal, Cartesian, coordinate system with respect to the Euclidean metric of \ $\mathbb{R}^n$. \     
Consider \ $A, B \subset \mathfrak{X}$. \  Then the outcomes of \ $A$ \ and \ $B$ \ are probabilistically independent if (6) holds.   
Does the probabilistic independence imply orthogonality? The answer, in general, is negative.   
However, these two concepts start having some common points if we consider the space \ $\mathfrak{X}$ \ having a large cardinality as is the case of interest in Statistical Mechanics.\\

To see how the relation between probabilistc independence and orthogonality occurs, one can start by moving the barycenter of \ $\Delta^{n-1}$ \ 
to the origin of \ $\mathbb{R}^n$ \ by a translation, or more generally an affine transformation.
It may be instructive, and more familiar from a physical viewpoint,  to consider the probabilities \ $p_i, \ i=1,\ldots, n$ \ as unit point masses placed on the 
respective axes at distances \ $p_i$ \ from the origin of the Cartesian coordinate system of \ $\mathbb{R}^n$ \ which we use. 
After the translation, the resulting simplex, let us call it \ $\Delta_\ast ^{n-1}$, \  
is the boundary of a convex body \ $\mathcal{K}$ \ whose center is now at the origin, 
and whose vertices have coordinates \ $p_i^\prime, \ i=1,2, \ldots,n$,  \  which means that
\begin{equation} 
                  \sum_{i=1}^n \ p_i^\prime  =  0
\end{equation}
Let us indicate the Euclidean inner product by \ $\langle \  , \  \rangle$. \ Let \ $e_1, \ldots , e_n$ \ indicate the unit vectors along the respective coordinate axes of the 
probabilities \ $p_1^\prime , \ldots, p_n^\prime$. \  Then, for vectors \ $u, v, w \in \mathbb{R}^n$ \ let us indicate the tensor product by \ $u \otimes v$, \ namely    
\begin{equation}
         (u\otimes v) (w) = \langle v, w \rangle u 
\end{equation}
In case the vector \ $u$ \ has unit length, then (19) implies that \ $u\otimes u$ \ is the orthogonal projection to the  linear subspace spanned by \ $u$. \   
Going back to the physical interpretation of the \ $p_i$, \ we can see that the ``inertia tensor'' of the simplex \ $\Delta_\ast ^{n-1}$ \  is not only diagonal, 
but even more so it is  the identity, namely
\begin{equation}
\sum_{i=1}^n \ p_i^\prime \ e_i\otimes e_i = 1_n 
\end{equation}
where \ $1_n$ \ indicates the identity map on \ $\mathbb{R}^n$. \ Then a theorem of F. John \cite{AGM1}, states that there is an ellipsoid of maximum volume contained 
within \ $\Delta_\ast ^{n-1}$ \ which is actually a Euclidean ball. Then, the inertia condition
(20) implies that such \ $e_i$ \ would behave like an orthonormal set since the inner product of any two vectors can be written as    
\begin{equation}
  \langle v, w  \rangle = \sum_{i=1}^n \ p_i^\prime \   \langle v, e_i \rangle \langle e_i, w \rangle
\end{equation}
for \ $v, \ w \in \mathbb{R}^n$. \\

  It turns out that John's theorem is valid for a general convex body \ $\mathcal{K}$ \ obeying conditions (18) and (20) and not just for a simplex like
  \ $\Delta_\ast^{n-1}.$ \ 
  In that case, one takes as unit vectors \ $f_i, \ i=1, 2, \ldots, n$ \ the normalizations of the vectors which are intersections \ $\Delta_\ast ^{n-1} \cap B_n$ \ where \ $B_n$ \ 
  indicates the maximum volume Euclidean ball. To be consistent with the formulation of the previous paragraph, the 
  coefficients \ $c_i^\prime, \ i=1,2,\ldots, n$ \ corresponding to \ $p_i^\prime$ \ in (18), and (20) are all taken to be positive.
  But then  it is no longer true that the unit vectors \ $f_i, \ i=1,2,\ldots, n$ \ 
  are orthogonal. The fact that it is possible to find an orthonormal basis \ $\{z_i, \ i=1,\ldots,n \}$ \ which is not ``too far'' from the basis \ $\{ f_i, \ i=1,2,\ldots, n \}$ \ of \
  $\mathbb{R}^n$ \ is the content of the Dvoretzky-Rogers lemma \cite{AGM1} which states that there is \ $\{ u_i, \ i=1,2,\ldots,n \} \subset \{ f_i, \ i=1,2, \ldots,n \}$, where 
  \ $u_i$ \ belongs to the linear span of \ $z_1, z_2, \ldots, z_i$, \ such that 
  \begin{equation}             
    \sqrt{\frac{n-i-1}{n}}  \leq \langle u_i, z_i \rangle \leq 1, \ \ \ \ \ \  i=1,2,\ldots n
  \end{equation}

Consider now two subsets \ $A, B \subset \mathfrak{X}$ \ and \ $A\cap B$ \ be their intersection as in (6). Assume that their intersection is a connected set, for simplicity. 
If it is not, then one can apply this reasoning to each connected component separately. The intersection of \ $A$ \ and \ $B$ \ will be a convex subset of the simplex \ $\Delta^{n-1}$. \ 
Then, this convex subset upon a possible affine transformation, satisfies the conditions of John's theorem, therefore it has a maximum volume inscribed ball for which the 
Dvoretzky-Rogers lemma applies. Hence, there are nearly orthonormal vectors forming a basis of \ $A\cap B$. \ We can extend this basis to an orthonormal basis of  
\ $A$ \ and \ $B$ \ respectively by a judicious choice of basis vectors for each one set, regardless of what choice we make in the other, since \ $A$ \ and \ $B$ \ 
do not intersect outside of \ $A\cap B$. \  
In essence we can extend the near orthonormal basis provided by the Dvoretzky-Rogers lemma to an almost orthonormal basis of \ $A$ \ and \ $B$. \ Hence the sets 
\ $A$ \ and \ $B$ \ appearing in (6) are almost orthogonal to \ $A\cap B$. \ The word ``almost'' appearing everywhere above  
disappears when one takes \ $n\rightarrow \infty$ \ as is assumed in many cases in 
Statistical Physics when one take the thermodynamic limit. Then  (22) implies that the majority of vectors \ $u_i$ \ and \ $z_i$ \ become collinear hence \ $A\cap B$ \ is the 
orthogonal intersection of sets \ $A$ \ and \ $B$ \ for the vast majority of their elements. Hence one can state, in this particular context, that probabilistic independence 
implies orthogonality, as we wanted to ascertain, answering question 2 of the previous Section. \\
              

\section{Conclusions and  discussion}

In this work we pointed out that the q-entropy composition property has the exact same form as the Pythagorean theorem in hyperbolic space. We justified the appearance of hyperbolicity in this composition by ascribing it to the simple relation between the q- and the $\kappa$- entropies where the latter are based on Lorentz trasformations, hence on Minkowski space-time, and by subsequently using the hyperboloid model of hyperbolic space to reach this conclusion. We justified the appearance of orthogonality through the use of the Dvoretzky-Rogers lemma applied to convex sets in high dimensional linear spaces.\\    

In a recent work \cite{LM}, motivated by the appearence of $\kappa$ distributions in astrophysical plasmas, the departure of the q-entropy composition 
from a simple addition seems to be considered as a drawback, as this is how  one may interpret the word ``defect'' in the title of that work. 
If one accepts the ordinary/usual addition as fundamental and all other forms of addition are compared to it, then the authors may be right. 
From the same perspective, if one accepts the BGS entropy and its composition as the ``reference definition'' and ``reference composition'' relation of entropies 
to which all other entropic forms  should be compared, and be considered deviations from, then the word ``defect'' may be applicable. \\

What the present work shows, in our opinion, is that the q-entropy, and we suspect many other entropies, are fundamentally different from the BGS one, 
as much as hyperbolic geometry is different from the Euclidean one. The usual addition is not ``better'' or ``more correct'' than a generalized addition such as 
the one expressed in (4), (9); they are just different. We should not consider  one as a variation of the other. 
They apply to different systems and serve different purposes. This is clearly along the lines of \cite{Tsallis} who repeatedly stresses that the q-entropy is not just a variation 
of the BGS entropy, as it might be inferred from its functional form, but a fundamentally different entropy which is applicable to different systems, 
and which happens to have as an appropriate limit the BGS entropy. In our work we show that the BGS and q-entropies have the same relation as Euclidean to hyperbolic
geometries; they are just different. \\

Given all this, the major, in our opinion, question remains; to which physical systems are the q- and $\kappa$- entropies applicable and what is the underlying 
dynamics that such  entropies describe at the macroscopic level? We believe that despite its difficulty, this may be a worthy subject for a future line of investigation.\\          



\end{document}